\documentclass[]{spie}  

\usepackage{amsmath,amsfonts,amssymb}
\usepackage{graphicx}
\usepackage[colorlinks=true, allcolors=blue]{hyperref}
\usepackage{xspace}
\usepackage{xcolor}

\newcommand{\red}{\textcolor{red}}

\def\m87{M87$^*$\xspace}
\def\sgra{Sgr~A$^*$\xspace}
\def\uas{$\mu$as\xspace}
\def\Glam{G$\lambda$\xspace}

\def\lsim{\mathrel{\raise.3ex\hbox{$<$\kern-.75em\lower1ex\hbox{$\sim$}}}}
\def\gsim{\mathrel{\raise.3ex\hbox{$>$\kern-.75em\lower1ex\hbox{$\sim$}}}}

\title{The Black Hole Explorer: Instrument System Overview}

\author[1]{Daniel P. Marrone}
    \affil[1]{Department of Astronomy and Steward Observatory, University of Arizona, 933 N Cherry Avenue, Tucson, AZ 85721, USA}
\author[2]{Janice Houston}
    \affil[2]{Center for Astrophysics $|$ Harvard \& Smithsonian, 60 Garden Street, Cambridge, MA 02138, USA}
\author[3,4,5]{Kazunori Akiyama}
    \affil[3]{Haystack Observatory, Massachusetts Institute of Technology, Westford, MA 01886, USA}
    \affil[4]{Mizusawa VLBI Observatory, National Astronomical Observatory of Japan, Iwate 023-0861, Japan}
    \affil[5]{Black Hole Initiative at Harvard University, 20 Garden Street, Cambridge, MA 02138, USA}
\author[6]{Bryan Bilyeu}
    \affil[6]{MIT Lincoln Laboratory, Lexington, MA 02421}
\author[6]{Don Boroson}
\author[2]{Paul Grimes}
\author[2]{Kari Haworth}
\author[7]{Robert Lehmensiek}
    \affil[7]{National Radio Astronomy Observatory, Charlottesville, VA 22903, USA} 
\author[8]{Eliad Peretz}
    \affil[8]{NASA Goddard Space Flight Center, Greenbelt, MD 20771, USA}
\author[2,5]{Hannah Rana}
\author[9]{Laura C. Sinclair}
    \affil[9]{National Institute of Standards and Technology, Boulder, Colorado 80305}
\author[7]{Sridharan Tirupati Kumara}
\author[2]{Ranjani Srinivasan}
\author[2]{Edward Tong}
\author[6]{Jade Wang}
\author[2]{Jonathan Weintroub}
\author[2,5]{Michael D. Johnson}

\authorinfo{Corresponding Author: Daniel P. Marrone (\url{dmarrone@arizona.edu})}

\begin{document} 
\maketitle

\begin{abstract}
The Black Hole Explorer (BHEX) is a space very-long-baseline interferometry (VLBI) mission concept that is currently under development. BHEX will study supermassive black holes at unprecedented resolution, isolating the signature of the ``photon ring'' --- light that has orbited the black hole before escaping --- to probe physics at the edge of the observable universe. It will also measure black hole spins, study the energy extraction and acceleration mechanisms for black hole jets, and characterize the black hole mass distribution. BHEX achieves high angular resolution by joining with ground-based millimeter-wavelength VLBI arrays, extending the size, and therefore improving the angular resolution of the earthbound telescopes. Here we discuss the science instrument concept for BHEX. The science instrument for BHEX is a dual-band, coherent receiver system for 80-320\,GHz, coupled to a 3.5-meter antenna. BHEX receiver front end will observe simultaneously with dual polarizations in two bands, one sampling 80-106\,GHz and one sampling 240-320\,GHz. An ultra-stable quartz oscillator provides the master frequency reference and ensures coherence for tens of seconds. To achieve the required sensitivity, the front end will instantaneously receive 32\,GHz of frequency bandwidth, which will be digitized to 64\,Gbits/sec of incompressible raw data. These data will be buffered and transmitted to the ground via laser data link, for correlation with data recorded simultaneously at radio telescopes on the ground. We describe the challenges associated with the instrument concept and the solutions that have been incorporated into the baseline design.
\end{abstract}


\section{Introduction}
\label{sec:intro}  

Supermassive black holes have been known for decades to be present in the centers of nearly all galaxies 
and intimately connected to their evolution. 
The mechanisms by which they grow, merge, and influence their host galaxies on scales far beyond the influence of their gravity, remain topics of intense study, 
as reconfirmed in the most recent Astronomy \& Astrophyics Decadal Survey\cite{astro2020}. 
These effects are anchored at event horizon scales, however, and therefore extremely difficult to observe directly. 
Despite harboring millions to billions of solar masses, the largest angular sizes of any supermassive black holes are $\sim$50~$\mu$as, which exceeds the angular resolution of nearly all astrophysical observations.
However, the groundbreaking event-horizon scale images made by the Event Horizon Telescope (EHT) collaboration over the past several years 
have provided our first glimpse of the universe at its innermost edges, around the black holes at the center of 
the elliptical galaxy M87 (\m87) and in our own Milky Way (\sgra) \cite{EHTC_M87_I,EHTC_M87_II,EHTC_M87_III,EHTC_M87_IV,EHTC_M87_V,EHTC_M87_VI,EHTC_M87_VII,EHTC_M87_VIII,EHTC_M87_IX,EHTC_SgrA_I,EHTC_SgrA_II,EHTC_SgrA_III,EHTC_SgrA_IV,EHTC_SgrA_V,EHTC_SgrA_VI,EHTC_SgrA_VII,EHTC_SgrA_VIII}.

The EHT observations are possible thanks to the technique of very-long-baseline interferometry (VLBI), in which radio telescopes are 
used to simultaneously record the radiation received from a black hole at positions separated by thousands of kilometers. 
The coherence of the radiation between every pair of points is computed from these recordings, providing measurements of the so-called ``visibility'' of the source.
Each visibility represents the Fourier transform of the sky emission, sampled on a spatial scale 
inversely proportional to the telescope separation. The combination of measurements 
made at telescopes up to 11,000~km apart on Earth has allowed the EHT team to produce images with an angular resolution of $\sim$20~$\mu$as, 
just sufficient to resolve these event horizons.

To advance black hole science in the next decade, it is imperative to push this these observations to yet higher resolution. 
Measurements on spatial scales of a few $\mu$as would be transformational for our understanding of black hole accretion, feedback, and 
demographics, particularly in the poorly known spin distribution of black holes. In particular, the black holes accessible to EHT-like observations 
are generally the common low-accretion rate black holes that pervade the universe, rather than the high-accretion rate objects most often 
observed for spin by, e.g., X-ray telescopes\cite{Reynolds_2014}. However, the angular resolution of Earthbound VLBI is limited by the ratio $\lambda/D$, which sets 
the angular resolution of optical systems. With the diameter of the Earth limiting us to $D\sim 10^4$~km and absorption in the atmosphere limiting radio observations 
to wavelengths of $\sim1$~mm and longer, the EHT already has reached the angular resolution limits achievable from the ground. To go further requires 
leaving the Earth.

The Black Hole Explorer (BHEX) is a NASA Small Explorer (SMEX) mission concept. As described in [\citenum{BHEX_Johnson_2024}], BHEX aims to explore the mass and spin, 
accretion processes, and jet launching effects of black holes to reveal how they grow, how they change their host galaxies, and to connect them to their 
electromagnetic emission to their newly observable multi-messenger signatures. 
BHEX follows previous space VLBI experiments \cite{Levy_1986,Linfield_1990}, and missions like HALCA/VSOP \cite{Hirabayashi_1998} and RadioAstron \cite{Kardashev_2013}, which 
observed at frequencies up to 22~GHz (1.3~cm wavelength). In order to have a clear view through through synchrotron emitting 
plasma around important black holes and to quickly image them at few \uas resolution, BHEX operates at much higher frequency, simultaneously observing 
at 100 and 300~GHz (3 and 1~mm wavelength). This significant increase in frequency places tighter restrictions on the capabilities of many elements of the system, 
as detailed below. In this paper we describe the instrumental design of BHEX, including many of the technical challenges and design trades that have been encountered 
thus far. The resulting instrument packages a set of well-proven space technologies into an unprecedented tool for understanding supermassive black holes.

\section{Driving Requirements}
\label{sec:reqs}
\begin{figure}
  \centering
  \centering    
    \includegraphics[width=0.85\textwidth]{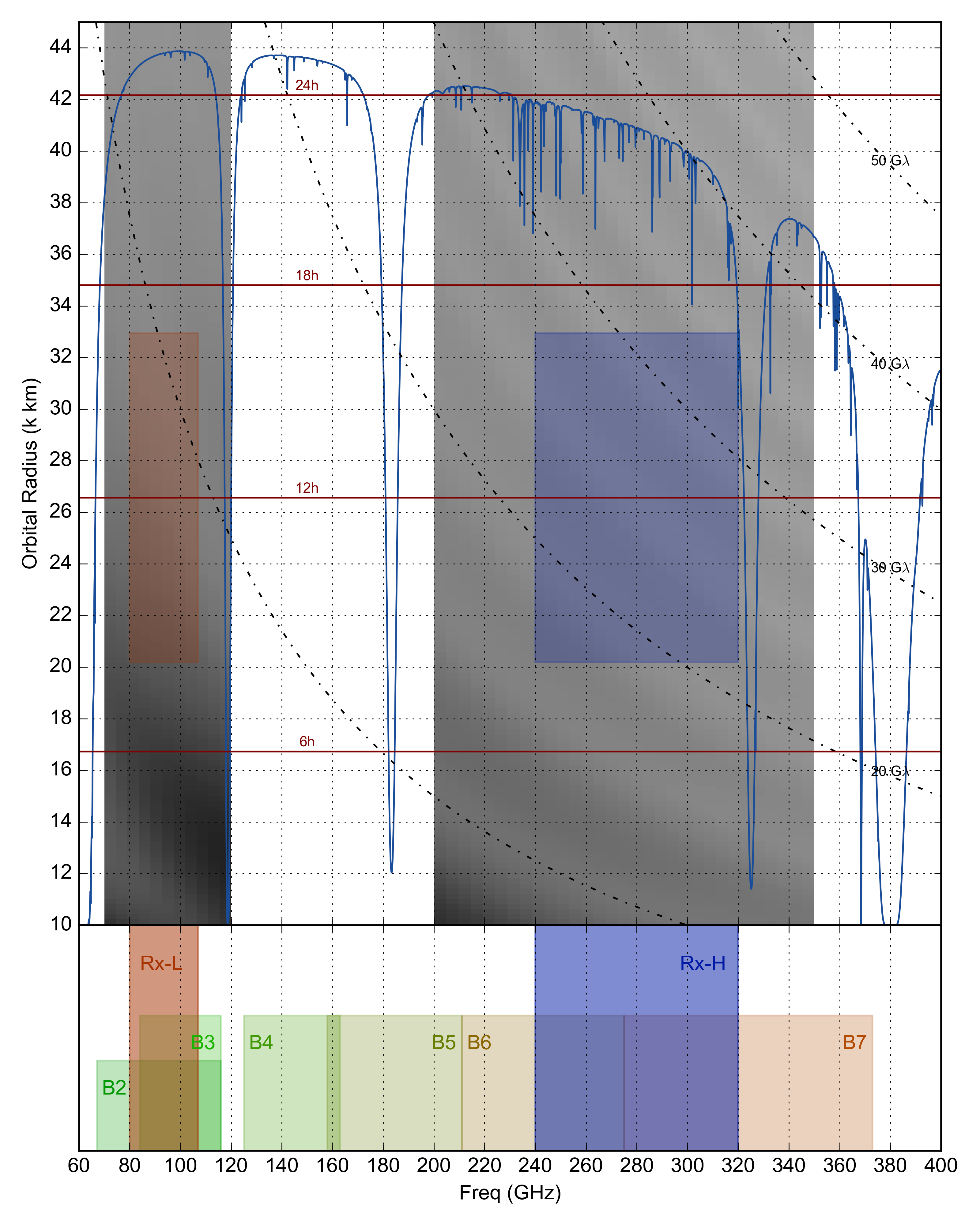}
  \caption{Constraints on the frequency and orbital distance for BHEX. The upper panel shows several effects as a function of frequency and orbital radius. The grey-scale shows the visibility of \m87 across the frequency-radius plane (for a baseline to the center of the Earth), with a square-root color stretch. The atmospheric transmission in median weather at the ALMA site is shown in blue, with 100\% at the top and 0\% at the bottom. Curves of constant baseline length, in billions of wavelengths (\Glam), are shown with dash-dotted lines. The radii corresponding to particular orbital periods, in sidereal hours, are shown in dark red. 
  For a chosen orbital radius, the range of available baseline lengths extends one Earth radius in either direction because ground sites are arrayed across the Earth, and so the frequency range and distance range covered by BHEX is represented by the red and blue shaded regions.
  The lower panel shows the selected frequency ranges for the two BHEX bands and the band frequency ranges for overlapping ALMA bands.
  }
  \label{f:orb}
\end{figure}

The scientific payload of BHEX must:\vspace{-1em}
\begin{enumerate}
\setlength{\itemsep}{0em}
\setlength{\parskip}{0em}
\item observe at a high enough frequency to see deep into black hole environments with few~\uas resolution 
\item with sufficient sensitivity to enable detection of the correlated flux density on these fine angular scales, 
\item receiving, digitizing, and transmitting the incident electric field 
\item while maintaining coherence with ground stations. 
\end{enumerate}
\vspace{-1em}We consider these individually below, and many of the constraints are visible within Figure~\ref{f:orb}.

The scientific goals of BHEX require {\bf(a)} that it see through dense synchrotron-emitting plasmas to study emission from just outside the event horizon of its primary targets. 
The synchrotron opacity falls with increasing frequency, suggesting that BHEX should operate at as high a frequency as is practical. 
Similarly, the interstellar scattering\cite{Psaltis_2018,Johnson_2018} that blurs the view of \sgra and the polarization modifying effects of propagation 
through magnetized plasma decrease with the square of the observing frequency, again favoring higher frequencies. 
Past EHT observations and analysis of simulated accretion flows\cite{BHEX_Lupsasca_2024} suggest that 
at frequencies above $\sim$200~GHz, BHEX should have a direct view of the event horizon regions of 
\sgra and \m87, and that the interstellar blurring of \sgra should be sub-dominant on \uas scales above 300~GHz \cite{Palumbo_2023}.
Simulations of these black holes\cite{BHEX_Lupsasca_2024} have suggested that the visibility signature of the photon ring should be measurable on baselines longer 
than $\sim$20~\Glam (resolution better than 10\uas), 
which corresponds to a projected telescope separation of at least 30,000~km at a frequency of 200~GHz ($\lambda$ = 1.5~mm)
or 20,000~km at 300~GHz ($\lambda$ = 1~mm). 

Observations at very high angular resolution require very high \textit{sensitivity} in order to detect correlated flux between the interferometer stations {\bf(b)}. 
The large separation between space and ground-based telescopes means that the baselines to space are sensitive to a high-pass-filtered version 
of the source structure, seeing only the very finest details in the emission and losing the smooth structures that contain most of the flux density. 
The details of the so-called visibility function, the visibility amplitude as a function of baseline length, depend on the precise source structure in complex 
sources. However, for the photon ring signature that is a core feature of the BHEX science goals robust predictions can be made for the visibility 
function, which depends primarily on the fraction of the total source flux density that is contained within the photon ring and the spatial extent of the photon ring. 
The first analytic predictions\red{\cite{Johnson_2020}} suggested a visibility of $V\sim 30\,{\rm mJy}\,(|{\bf u}|/10\,{\rm G}\lambda)^{-3/2}(S_{\rm tot}/1\,{\rm Jy})$, for baseline 
length $|{\bf u}|$ in wavelengths and total flux source flux density $S_{\rm tot}$. These estimates, and subsequent calculations with GRMHD simulations and 
similar synthetic image structures\red{\cite{Cardenas_2023}}, suggest a visibility of 5-10~mJy on baselines of 20-30~\Glam at a frequency near 230~GHz, 
where the EHT observes. The BHEX sensitivity must be such that a space-ground baseline can detect such a weak signal within the visibility measurement time.

Important technical requirements {\bf (c)} flow from this sensitivity requirement, as will be discussed in more detail below. The VLBI technique requires that BHEX use phase preserving technologies such as mixers and amplifiers to capture its signals, which have practical and quantum mechanical limits on their sensitivity. 
The sensitivity of a space-ground baseline is inversely proportional to the square-root of the spectral bandwidth that can be sampled, so this bandwidth must be 
maximized. However, a competing limitation is imposed by the fact that the data cannot be averaged on orbit, only after correlation with the signal received at other 
telescopes, and so digitization of a total analog bandwidth $\Delta\nu$ at a resolution of $B$ bits implies the production of data at a rate 
$R = 2\,\Delta\nu\,B$~bits per second, where the leading factor of two comes from the need to Nyquist sample the data stream. 
Consider, for scale, the current EHT receiver systems. The nominal receiver configuration instantaneously ingests 
$\Delta\nu$ = 16~GHz of spectral bandwidth, which is digitized and recorded at $B$ = 2 bits of resolution, for a total 
data rate of 64~Gbits per second (Gbps). For comparison, the data rate transmitted to the ground in real time by RadioAstron was 500$\times$ smaller. 
Transmitting such wide data bandwidths to the ground has been implausible until recently.

The data transmitted down from the BHEX satellite must also retain a detectable correlation with the light received at ground stations {\bf (d)}. 
This requires averaging in frequency and in time, with the sensitivity improving like the square root of the integration time $\tau$. 
However, this is only possible if the BHEX data faithfully capture the oscillations of the incident light, rather than blurring them 
by mixing with an imperfect reference oscillator or through the stochastic change in atmospheric path length above the ground station. 
Phase noise in the reference oscillator (at $\nu_{\rm ref}$ = 10 to 100~MHz) will be amplified by multiplication up to the observing frequency, 
with the phase noise growing like the square of this multiplication ratio, and therefore the square of the observing frequency. 
Atmospheric path variations are caused by turbulence in the distribution of water in the troposphere, and create 
arrival time changes that are constant with frequency for the frequencies of interest, which means that they introduce phase errors 
that grow linearly with frequency. Both of these effects argue for observing at the lowest frequency possible. 
Given the astrophysical constraints {\bf(a)} described above, BHEX is designed to use a lower-frequency band to measure and remove 
errors introduced by the reference oscillator and atmosphere, allowing longer averaging times for the detection of the visibility at $\sim$300~GHz.

\section{Instrument Concept}
\begin{figure}
  \centering
  \centering    
    \includegraphics[width=\textwidth]{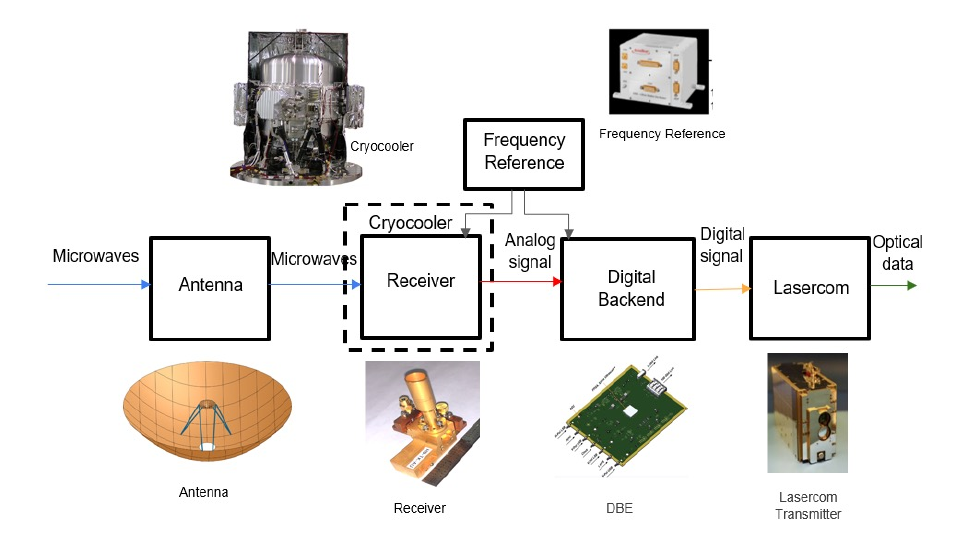}
  \caption{Schematic diagram of the BHEX instrument subsystems and their interrelation. 
  }
  \label{f:bd}
\end{figure}

The BHEX science instrument is a radio telescope and (heterodyne) receiver system, similar to the radio frequency (RF) communications technologies used in nearly all space missions. 
As shown in the block diagram in Figure~\ref{f:bd}, it consists of 1) an antenna, 2) dual-band radio receiver, 3) cryocooler, 4) digital back end, 5) frequency reference, and 6) optical data transmission subsystem. 
The 3.5-meter antenna will be made of metallized carbon fiber. 
The BHEX receiver will observe simultaneously in two bands, one around 100 GHz (the low-band receiver, RxL) and one around 300 GHz (high band, RxH), sampling both polarizations in each band. To achieve the necessary sensitivity, this receiver will be cooled by a Stirling/J-T space cryocooler. 
The total analog bandwidth produced by the receiver is 32 GHz, which will be digitized at 1-bit resolution for a total data rate of 64~Gbps. The data will be transmitted to the ground via the optical data transmission subsystem.
An ultra-stable quartz oscillator (USO) provides the master frequency reference and ensures coherence for tens of seconds at 100~GHz. 

The BHEX instrument design leverages space heritage. The antenna design is adapted from current development work for an Earth-observing satellite and is based on the telescope for the Planck mission. Superconductor-insulator-superconductor (SIS) mixers like those used in the high-band receiver have previously flown on the Herschel mission\cite{deGrauuw_2010} and cryogenic 100~GHz high electron mobility transistor (HEMT) amplifiers like those in the low-band receiver flew in the Wilkinson Microwave Anisotropy Probe\cite{Bennett_2003}.
A suitable cryocooler\cite{XRISM_cryo} is currently operating successfully in the XRISM mission. The digital backend is based upon state-of-the art designs which must operate in the harsh radio telescope environments (e.g., South Pole, 17,000-foot altitude in Chile), upgraded to space-grade components. The frequency reference recently launched on a mission to Jupiter\cite{juice}. The ultra-high bandwidth optical data transmission has been successfully demonstrated from a cubesat mission\cite{Schieler_2023}

\subsection{Antenna}
The 3.5~m antenna will be made of metallized carbon fiber. As the space-based component of a space-ground interferometer, the sensitivity of each space-ground baseline (formed by correlating the electric fields received
at BHEX and a ground-based observatory) is proportional to the geometric mean of the space and ground antenna collecting areas, allowing BHEX to be implemented with an aperture suitable for launch in available fairings. 

The optical design of the system is described in [\citenum{BHEX_Sridharan_2024}] and [\citenum{BHEX_IEEE_Lehmensiek_2024}]. The telescope is currently an on-axis Gregorian design, with an offset between the axis of revolution that defines the telescope surface and the common axis of symmetry for the parabola and ellipse that respectively define the primary and secondary mirrors. 
This produces a ring focus between the primary and secondary mirrors, but allows for $\sim$90\% aperture efficiency, maximizing the sensitivity.

\subsection{Receiver}
The requirements described in section~\ref{sec:reqs} 
imply the need for simultaneous observations in two receiver bands. The lower-frequency band provides additional science, but is critically important to simplify ``fringe finding'' -- the process of identifying correlation between the light recorded on orbit and at ground-based stations -- 
and to remove atmospheric and timing-based errors from the higher-frequency band. 
At 100~GHz, decoherence from the atmosphere and the frequency references on the ground and in space is substantially less important. Moreover, because the 
baseline is a factor of 3 shorter in units of the observing wavelength, the corresponding angular resolution is 3$\times$ lower and the visibility is significantly greater 
for many sources (many BHEX targets also have higher total flux density at 100~GHz). 

The frequency range of the high-band receiver is subject to the constraints discussed in Sec.~\ref{sec:reqs}, 
but also restrictions imposed by the transmission of the Earth's atmosphere. This is because 
BHEX will operate as a hybrid observatory in concert with ground-based stations that must be able to observe at the same frequency.
Water vapor is the primary limiting component for observations above $\sim$120~GHz, with strong lines at 183~GHz and 325~GHz providing important 
practical limits. Even at the highest and driest observatories on Earth, such as the Atacama Large Millimeter/Submillimeter Array (ALMA) in Chile, 
the impact of water is such that observations at frequencies above $\sim$320~GHz are confined to the best $\sim$40\% of the site weather. 
We therefore adopt 320~GHz as the upper frequency of the band. 
For a semi-synchronous (12 sidereal hour) orbit, which has significant advantages 
for coordinated observations with ground stations, this observing frequency provides a maximum baseline of $\sim$35~\Glam, or 6~\uas resolution, meeting requirement {\bf (a)}. 
The high band uses SIS mixers, which are the most sensitive technology available at these frequencies. 
A fractional bandwidth ($\nu_{\rm max}/\nu_{\rm min}$) of 4/3 is readily achieved for SIS mixers,
corresponding to a frequency range of 240 to 320~GHz. 

The low-band receiver frequency range is driven by its role as a fringe-finding band. We will employ the 
``frequency phase transfer'' (FPT) technique to remove visibility phase variations that are common to both bands, namely those due to atmospheric perturbations and clock variations.\cite{Rioja_2020}
These effects create phase erorrs that scale linearly with frequency. The low-band frequency range is set to allow the bands to be used at an integer frequency ratio so that any ambiguity in the wrapping of 100~GHz phases does not 
introduce ambiguity in the phase correction at 300~GHz. A weak frequency comb will be continuously injected to both bands to enable tracking of 
any instrumental phase drifts between or within bands that could corrupt the FPT.
Scaling the high band frequency down by a factor of three, 
we find a low-band frequency range of 80 to 107~GHz, which 
fits nicely between strong atmospheric features from molecular oxygen below 65~GHz and above 118~GHz. 

The two receiver bands each produce 8~GHz of analog bandwidth per polarization, for a total of 32~GHz. The low band is a single-sideband system, so only 8~GHz of sky frequency is covered. The high band is a double-sideband system with an intermediate frequency range of 4-12~GHz.
This introduces a noise penalty, but provides twice as much frequency coverage per transmission bandwidth, which improves the range of baseline lengths (in wavelength units) instantaneously sampled by BHEX. The covered frequencies extend across 24~GHz (12~GHz below the local oscillator frequency to 12~GHz above, missing the central 8~GHz), a perfect match to the 8~GHz low-band coverage, after accounting for the factor of three frequency ratio.

A more detailed discussion of the receiver design can be found in [\citenum{BHEX_Tong_2024}].

\subsection{Frequency Reference}
A sufficiently stable frequency reference ensures coherence with the ground on timescales, 100's of $\mu$s - 10's of s, in order to find correlation (``fringes,'' in the interferometry jargon).  Unlike the earlier RadioAstron Mission\cite{Kardashev_2013}, the high observing frequency of 345 GHz, places more stringent requirements on the frequency reference. Despite this, there are a number of different approaches, both optical and microwave, to generating the required 10 or 100~MHz output with sufficiently low phase noise to support the performance requirement. The driving factor in choice of frequency reference then becomes the combination of performance with minimization of the SWaP-envelope. (See [\citenum{BHEX_Peretz_2024}] for trade-offs associated with differing approaches.) Additionally, a coarser time reference provides a measure of the drift rate of that frequency reference to better than 1 ps/sec and allows for the narrowing of the search window for fringes. 

A primary challenge for all VLBI observations is the need to preserve coherence between telescopes. A geographically distributed telescope array generally cannot share a single frequency reference, and any imperfections in the independent references at each site will get encoded into the downconverted and sampled data streams recorded at each site. Time and frequency offsets are corrected at the correlation stage, but fast random errors (e.g., frequency/phase jitter) reduce the correlation between sites, falsely implying that the source visibility itself is weaker than it actually is. For ground-based telescopes, a demonstrated solution for VLBI observations up to 350~GHz is to use an active hydrogen maser as the frequency reference. These devices are massive and delicate, but have been used in space. The RadioAstron space VLBI mission included a maser\cite{Kardashev_2013}, though its performance requirements were not as stringent as those of BHEX because its maximum operating frequency was more than ten times lower.

A maser is a difficult fit for a SMEX mission like BHEX. Fortunately, the ESA Jupiter Icy Moons Explorer (JUICE) mission\cite{juice} recently launched with a state-of-the-art ultra-stable oscillator\cite{Accubeat_press, NIST_disclaimer} (USO) that is small, low-power, with stability comparable to ground-based masers for timescales of 1 second and shorter. This quartz crystal USO, perhaps with small improvements in its performance on timescales of a few seconds, is a viable reference for the BHEX mission. Other reference technologies have also arisen in the last few years, which may be of interest for BHEX. A commercial company recently demonstrated a highly stable oscillator based on a modulated laser locked to a transition of the iodine module designed for operation in poorly controlled environmental conditions, such as on board a ship at sea\cite{roslund_optical_2024}. Such an iodine clock, if packaged for space operation and optimized for lower power consumption, would also be a viable reference for the BHEX mission. 

\subsection{Cryocooler}
To achieve the sensitivity requirements of the BHEX instrument, portions of the receiver system must be cooled to cryogenic temperatures. The SIS mixers, in particular, function only when cooled to below the critical temperature of their superconducting layers ($T_c\sim9$~K in the case of Nb), and achieve their best performance at temperatures at or below $T_c/2$, or 4.5~K. Similarly, though they do not require cooling, the noise performance of the HEMT amplifiers in the low-band receiver is substantially improved by operating them at very low temperature. 

Fortunately, there has been an increasing need for compact, efficient, and closed-cycle cryocoolers that can achieve $\sim$4~K base temperatures in space applications \cite{ranacryo}. While some astrophysics missions have been cooled with liquid cryogens to 4~K or below (e.g., Herschel \cite{herschelcryocooling}), closed-cycle coolers with base temperatures of 4-6~K have flown on missions including Planck, JWST (MIRI), and XRISM \cite{Morgante_2009,JWST_MIRI,XRISM_cryo}. These operate with a combination of Stirling and Joule-Thomson (J-T) cooling. BHEX aims to leverage existing cryogenic technology, where the majority of the key cryogenic components within these technologies, such as Stirling and J-T coolers, boast established spaceflight heritage.

A number of design trade-offs are necessary for the instrument operation and the cryogenic system operation to ensure that both the scientific goals are met and the cryocooling system can be appropriately scaled down to fit within the constraints of a SMEX mission. Operation of the SIS mixers and low noise amplifiers sets the cooling power required from the cryocooler, which drives the power needed from the spacecraft for its operation. Though intermittent operation of the cryocooler could lessen the power burden, continuous operation is needed to maintain receiver stability. This requirement is beneficial, however, as it also serves to reduce the risk of contamination (the primary cause of cryocooler failure), and removes the need to wait for cool-down and temperature stabilization when initiating each observation cycle. The art of designing the cryogenic system for BHEX lies within striking a balance between mass, power, and cost reduction for SMEX constraints while maintaining the desired cooling performance and operation, and reducing failure risks.

The BHEX receiver subsystem demands approximately 10 and 125~mW at 4.5~K and 20~K, respectively, to cool the amplifiers of the low- and high-band receivers. This represents a combination active power sources, dominated by the power dissipation in the 100~GHz HEMT amplifiers (20~K) and the high-band IF amplifiers at 4.5~K, and parasitic radiative and conductive heat loads. There are multiple cryocoolers capable of removing this amount of heat, and they can be expected to consume 250~W or more. Optimizing the choice of cryocooler is an important ongoing trade study in the BHEX mission. For additional details on the space heritage, trade space, and design of the BHEX cryocooler, see [\citenum{BHEX_Rana_2024}].

\subsection{Backend}
The receiver generates four 4-12~GHz IF signal streams representing the received astronomical signal, a total of 32~GHz of analog bandwidth. As described in detail in [\citenum{BHEX_Srinivasan_2024}], the role of the backend subsystem is to translate these analog signals to Nyquist-sampled digital signals that can be transmitted to the ground. 
The backend analog processing components divide the 4-12 GHz signals into two sub-bands, each of which is 4~GHz wide. This processing also includes 
band-defining filters to eliminate aliasing of unwanted frequencies when they are digitized, since the receiver itself does not intentionally control the IF bandwidth. 
The eight sub-bands are amplified to a level suitable for digitization, and a power detector records the power in each sub-band because this information is lost in the 1-bit digitization.

The heart of the backend is the digitization system, in which each of the 4GHz-wide analog data streams is converted to a bitstream for downlink. 
This system follows from a line of such ``digital backends'' built for early EHT experiments, the EHT itself, and for wider bandwidth successors. 
For the BHEX application, the core components must be selected from the limited number of products designed for use in space. Fortunately, 
there are several commercial radiation-tolerant analog to digital converters (ADCs, or samplers) available, with sufficient RF bandwidth and sampling rates to ingest the BHEX analog sub-bands and represent them at 10 or more bits of precision.
The digitized samples will be processed by the other major component, a space-grade FPGA. In the current design this is nominally an AMD/Xilinx Versal FPGA. 
Within the FPGA, the high-resolution samples will be reduced to 1-bit resolution for downlink and packaged with header information, before relaying to the laser downlink.

The use of 1-bit resolution is a balance between sensitivity and transmission bandwidth. For many years, the standard approach to digitization for ground-based interferometric correlation under the constraints of limited signal transmission bandwidth was to digitize at 2-bit resolution. This provides 88.3\% of the signal-to-noise ratio present in the data, 
whereas 1-bit would provide just 63.7\% of the signal-to-noise (see, e.g., [\citenum{TMS}]), though with half the data rate. It is notable that for cases where the receiver can provide wider bandwidth before digitization, roughly the same SNR is expected for 2-bit sampling of bandwidth $\Delta\nu$ and 1-bit sampling of $2\Delta\nu$, both of which have a digital data rate of $4\Delta\nu$. For BHEX, the primary limitation on data transmission rates is from space to ground, transport of recorded bits across the surface of the Earth is relatively easy through the internet or shipped hard drives. We have therefore chosen to use 1-bit resolution in space and 2-bit on the ground, which, when correlated, will yield 75.0\% of the SNR. Compared to digitizing half as much analog bandwidth in space (and ground) at 2-bit resolution, this is a sensitivity gain of 20\% ($0.750\times\sqrt{2}/0.883$). 

\subsection{Optical Downlink}
With a 64~Gbps total data rate, a standard radio-frequency downlink would be extremely challenging for BHEX. Space optical communication has long been recognized as capable of supporting higher data rates with smaller apertures and lower on-orbit power consumption, and is therefore a key enabling technology for this mission. 

In 2022-2023, the TBIRD (TeraByte Infrared Delivery) mission demonstrated a 200 Gbps downlink from a cubesat in low-earth orbit (LEO)\cite{Riesing_2023,Schieler_2023}. As described elsewhere in these proceedings\cite{BHEX_Wang_2024}, the BHEX optical modem is derived from the TBIRD mission. The beam directing optics will make use of a larger aperture, which is required to provide $\sim$100 Gbps from a higher orbit, but suitable optical terminals have been demonstrated already and are available from commercial sources.

\section{Summary}
We have outlined the instrument design considerations that guide us in designing the science instrument for the BHEX mission concept. The overarching mission concept is described in an accompanying paper in these proceedings\cite{BHEX_Johnson_2024}. The science inputs have enabled us to construct a high-level design for a space instrument that is compatible with the constraints expected for an Explorer mission, and we have described many design choices and implementation details here. A summary of the expected parameters of the system are included in Table~\ref{t:summary}.  Additional details can be found in these and other proceedings for many of the subsystems, including the antenna\cite{BHEX_Sridharan_2024,BHEX_IEEE_Lehmensiek_2024}, the receiver\cite{BHEX_Tong_2024}, the backend\cite{BHEX_Srinivasan_2024}, the cryocooler\cite{BHEX_Rana_2024}, and the optical downlink\cite{BHEX_Wang_2024}.

\begin{table}[ht]
\caption{BHEX Mission Characteristics.} 
\label{t:summary} 
\begin{center}       
\begin{tabular}{|l|l|l|} 
\hline
\rule[-1ex]{0pt}{3.5ex} {\bf Antenna} & \\
\rule[-1ex]{0pt}{3.5ex} \quad Diameter & $3.5$-m\\
\rule[-1ex]{0pt}{3.5ex} \quad Optical Configuration & Axially Symmetric, Ring Focus \\
\rule[-1ex]{0pt}{3.5ex} \quad Surface RMS & 40\,$\mu$m\\
\hline
\rule[-1ex]{0pt}{3.5ex} {\bf Receiver} & \\
\rule[-1ex]{0pt}{3.5ex} \quad Number of Bands/Polarizations & 2 Bands (Co-pointed), 2 Polarizations \\
\rule[-1ex]{0pt}{3.5ex} \quad High Band (240-320 GHz) & \\
\rule[-1ex]{0pt}{3.5ex} \quad \quad Front End Technology & \quad SIS Mixer, cooled to 4.5~K \\
\rule[-1ex]{0pt}{3.5ex} \quad \quad IF Configuration & \quad 4-12 GHz, Double Sideband \\
\rule[-1ex]{0pt}{3.5ex} \quad \quad System noise temperature (DSB) & \quad 23-30 K\\
\rule[-1ex]{0pt}{3.5ex} \quad \quad SEFD at 240\,GHz & \quad 16,400~Jy \\
\rule[-1ex]{0pt}{3.5ex} \quad \quad SEFD at 320\,GHz & \quad 23,600~Jy  \\
\rule[-1ex]{0pt}{3.5ex} \quad Low Band (80-107 GHz) & \\
\rule[-1ex]{0pt}{3.5ex} \quad \quad Front End Technology & \quad InP HEMT Amplifier, cooled to 20~K \\
\rule[-1ex]{0pt}{3.5ex} \quad \quad IF Configuration & \quad 4-12 GHz, Single Sideband \\
\rule[-1ex]{0pt}{3.5ex} \quad \quad System noise temperature (SSB) & \quad 45~K \\
\rule[-1ex]{0pt}{3.5ex} \quad \quad SEFD at 80\,GHz & \quad 14{,}700~Jy \\
\hline
\rule[-1ex]{0pt}{3.5ex}  {\bf Science Data} &  \\
\rule[-1ex]{0pt}{3.5ex} \quad Quantization & 1-bit \\
\rule[-1ex]{0pt}{3.5ex} \quad Data Rate & 64 Gbps \\
\rule[-1ex]{0pt}{3.5ex} \quad Downlink Technology & Laser Transmission \\
\hline
\rule[-1ex]{0pt}{3.5ex} {\bf Payload Resources} & \\
\rule[-1ex]{0pt}{3.5ex} \quad Mass & 200~kg (CBE) \\
\rule[-1ex]{0pt}{3.5ex} \quad Power & 670~W (CBE) \\
\hline
\rule[-1ex]{0pt}{3.5ex} {\bf Mission Lifetime} & $2\,{\rm yr}$ \\
\hline 
\end{tabular}
\end{center}
\end{table}

\acknowledgments 

Technical and concept studies for BHEX have been supported by the Smithsonian Astrophysical Observatory, by the Internal Research and Development (IRAD) program at NASA Goddard Space Flight Center, by the University of Arizona Space Institute, and by the ULVAC-Hayashi Seed Fund from the MIT-Japan Program at MIT International Science and Technology Initiatives (MISTI). We acknowledge financial support from the Brinson Foundation, the Gordon and Betty Moore Foundation (GBMF-10423), and the National Science Foundation (AST-2307887, AST-2107681, AST-1935980, and AST-2034306). This work was supported by the Black Hole Initiative at Harvard University, which is funded by grants from the John Templeton Foundation and the Gordon and Betty Moore Foundation to Harvard University. BHEX is funded in part by generous support from Mr. Michael Tuteur and Amy Tuteur, MD. BHEX is supported by initial funding from Fred Ehrsam.

\bibliography{report,bhexspiepapers} 

\begin{thebibliography}{10}
\bibitem{NIST_disclaimer}
``Certain equipment or instruments are identified in this paper in order to
  specify the experimental procedure adequately. such identification is not
  intended to imply recommendation or endorsement of any product by nist, nor
  is it intended to imply that the equipment identified are necessarily the
  best available for the purpose..''
\end{thebibliography}
\bibliographystyle{spiebib} 

\end{document}